\documentclass[superscriptaddress,amsmath,amssymb,aps,prl,twocolumn,floatfix]{revtex4-2}

\usepackage[utf8]{inputenc}
\usepackage{todonotes}
\usepackage{xcolor}
\usepackage[colorlinks, linkcolor=blue, citecolor=blue, urlcolor=blue]{hyperref}
\usepackage{graphicx}
\usepackage{physics}
\usepackage{siunitx}
\usepackage{booktabs}
\usepackage{multirow}
\usepackage{longtable}

\usepackage[normalem]{ulem}

\newcommand{\mi}{\mathrm{i}}
\renewcommand{\vec}[1]{\mathbf{#1}}
\renewcommand{\d}{\mathrm{d}}

\begin{document}

\title{Collective Dissipative Molecule Formation in a Cavity}

\author{David Wellnitz}
\affiliation{ISIS (UMR 7006) and icFRC, University of Strasbourg and CNRS, 67000 Strasbourg, France}
\affiliation{IPCMS(UMR 7504), University of Strasbourg and CNRS, 67000 Strasbourg, France}

\author{Stefan Sch\"utz}
\affiliation{ISIS (UMR 7006) and icFRC, University of Strasbourg and CNRS, 67000 Strasbourg, France}
\affiliation{IPCMS(UMR 7504), University of Strasbourg and CNRS, 67000 Strasbourg, France}

\author{Shannon Whitlock}
\affiliation{ISIS (UMR 7006) and icFRC, University of Strasbourg and CNRS, 67000 Strasbourg, France}

\author{Johannes Schachenmayer}
\thanks{schachenmayer@unistra.fr}
\affiliation{ISIS (UMR 7006) and icFRC, University of Strasbourg and CNRS, 67000 Strasbourg, France}
\affiliation{IPCMS(UMR 7504), University of Strasbourg and CNRS, 67000 Strasbourg, France}

\author{Guido Pupillo}
\thanks{pupillo@unistra.fr}
\affiliation{ISIS (UMR 7006) and icFRC, University of Strasbourg and CNRS, 67000 Strasbourg, France}
\affiliation{Institut Universitaire de France (IUF), 75000 Paris, France}
	
\begin{abstract}
We propose a mechanism to realize high-yield molecular formation from ultracold atoms. Atom pairs are continuously excited by a laser, and a collective decay into the molecular ground state is  induced by a coupling to a lossy cavity mode.
Using a combination of analytical and numerical techniques, we demonstrate that the molecular yield can be improved by simply increasing the number of atoms, and can overcome efficiencies of state-of-the-art association schemes. We discuss realistic experimental setups for diatomic polar and nonpolar molecules, opening up collective light matter interactions as a tool for quantum state engineering, enhanced molecule formation, collective dynamics, and cavity mediated chemistry.
\end{abstract}

\date{\today}
	
\maketitle

There is considerable interest in preparing and manipulating ultracold ensembles of molecules for quantum simulations, metrology and the study of chemical reactions in the ultracold regime~\cite{Doyle_Quo_2004,Carr_Cold_2009,Quemener_Ultracold_2012,Moses_New_2017,Bohn_Cold_2017}. Diatomic molecules in their electronic and rovibrational ground state are routinely produced using the coherent stimulated Raman adiabatic passage (STIRAP) technique~\cite{Danzl_Ultracold_2010,Takekoshi_Ultracold_2014,Lang_Ultracold_2008,Moses_Creation_2015,DeMarco_Degenerate_2019}. Alternatively, continuous formation of ground state molecules can be realized by photoassociation via a weakly bound excited molecular state~\cite{jones_ultracold_2006,Deiglmayr_Formation_2008,bellos_formation_2011,Zabawa_Formation_2011, ulmanis_ultracold_2012,bruzewicz_continuous_2014}. While more sophisticated methods such as photoassociation followed by pulsed population transfer~\cite{Sage_Optical_2005} or re-pumping of vibrationally excited molecules~\cite{Viteau_Optical_2008,passagem_continuous_2019} have been experimentally demonstrated, efficiencies of ground state molecular formation are usually lower than those achieved with STIRAP and without rotational state selectivity. It has recently been proposed that these efficiencies can be increased by strengthening light-molecule coupling rates to ground-state transitions using a cavity~\cite{Kampschulte_Cavity_2018} or a photonic waveguide~\cite{Perez_Ultracold_2017}.
Common to all these schemes is the use of formation processes based on single molecules. 

\begin{figure}
	\centering
	\includegraphics[width=\columnwidth]{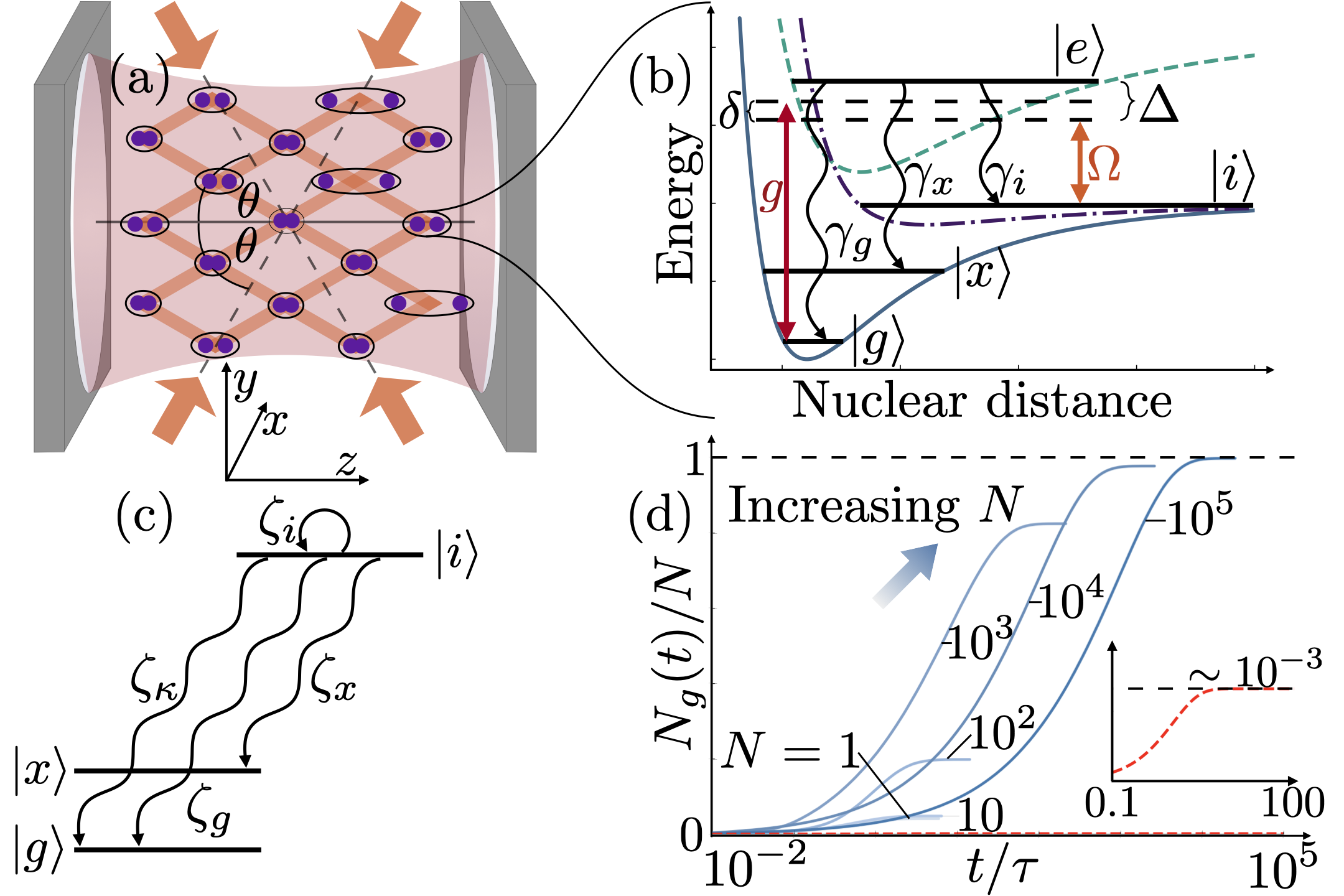}
	\caption{Basic setup for collective dissipative molecule formation. (a) Feshbach molecules are trapped in a lattice inside a cavity and brought into deeply bound states by photoassociation. An angle $\theta$ between the lattice laser beams and the cavity ($z$) axis ensures mode matching. (b) Scheme of energy levels and their coupling for a single molecule. For RbCs the potential energy curves can be identified with the ground state potential $X^1\Sigma^+$ (continuous line; dissociates to $5s+6s$), triplet ground state potential $ a^3\Sigma^+ $ (dash-dotted line; dissociates to $ 5s + 6s $) and excited state potential $ (A^1\Sigma^+ - b^3\Pi)0^+ $ (dashed line; dissociates to $ 5s + 6p $). A molecule prepared in a Feshbach state $ \ket{i} $ is laser excited (coupling strength $ \Omega $, detuning $ \Delta $) into the excited state $ \ket{e} $, which can decay back into $ \ket{i} $, the rovibrational ground state $ \ket{g} $ or any other state (bound or not), here collectively called $ \ket{x} $. (c) Energy levels of a single molecule after adiabatic elimination of the cavity and excited state with decay rates $\zeta_\alpha$ ($\alpha$=$\kappa$, $i$, $x$, $g$). (d) Evolution of the target ground state molecular fraction $N_g/N$ as a function of rescaled time $t$ (see text), for different $1\leq N \leq 10^5$ (log-scale). The red dashed line in the Inset indicates the results without cavity.}
	\label{fig:setup}
\end{figure}

Here, we propose a mechanism to exploit collective effects to perform continuous high-yield molecular formation from ultracold atoms in a cavity. Our scheme is based on photoassociation to a collective excited bound-state followed by superradiant-type decay induced by the cavity to the molecular ground state. 
We consider the regime of large dissipation with negligible number of cavity photons and electronic excitations, and derive an effective master equation for the internal dynamics of $N$ atom pairs. We show that (i) a continuous laser gives rise to enhancement of the fraction of ground state molecules $N_g^{\infty}/N \sim [1- \log(N)/(NC)]$ approaching 1, with $NC$ the collective cooperativity; and (ii) a chirped laser pulse that matches the time-varying excited molecular polariton energies can lead to a final molecular yield  $N_g^{\infty}/N \sim 1- \Gamma/\kappa$, with $\Gamma$ the excited state linewidth and $\kappa$ the cavity linewidth. The two schemes are most useful for weak and strong cavity couplings, respectively, for which we provide concrete examples. With scheme (i) collective effects always increase the molecular yields at the cost of decreased transfer rates. In contrast, scheme (ii) cannot be directly compared to the single particle scenario with a cavity, but always has higher yields than single-particle photo-association without a cavity. 
Both schemes can serve as alternatives to STIRAP that relax the requirement for expensive, narrow linewidth, phase-coherent lasers~\cite{vitanov2017stimulated}, and offer a natural way to continuously populate a molecular lattice coupled to a cavity. More broadly, this work exemplifies the opportunities for state engineering using collective effects in the presence of strong dissipation.

\medskip

We consider a setup consisting of $N$ identical pairs of atoms and a single mode cavity. The external dynamics of each pair is assumed to be frozen, e.g., by confining to an optical lattice potential. We model each atom pair as a four-level system with states $\ket{i}_n, \ket{e}_n, \ket{g}_n$ and $\ket{x}_n$, $1\leq n \leq N$. The first three states correspond to a two-atom initial state (e.g., a low-energy scattering state or pre-formed Feshbach molecular bound state), a molecular excited state and the absolute electronic and rovibrational molecular ground state, respectively [see Fig.~\ref{fig:setup}(b)]. The fourth level $\ket{x}_n$ represents a set of arbitrary excited molecular (e.g. vibrationally or rotationally excited) or free particle states, whose population we want to avoid. The dynamics of the system's density matrix $\hat \rho$ is governed by the master equation $\partial_t \hat{\rho} = -\mi [\hat{H}, \hat{\rho}]  + \mathcal{D} \hat{\rho}$,
with $\hat H = \hat H_{LA} + \hat H_C + \hat H_0$ the system Hamiltonian and ($\hbar=1$)
\begin{align}
	\label{eq:Ham_L}
	\hat{H}_{LA} &= \Omega \sqrt{N} \left(\hat S_{i e} + \hat S_{e i} \right) 
	\\
	\label{eq:Ham_C}
	\hat H_C &=  g \sqrt{N} \left( \hat  a^\dag \hat S_{g e} + \hat S_{e g} \hat  a \right).
\end{align}
Here, $\hat H_{LA}$ and $\hat H_{C}$ represent the coupling of the transition dipole moments of the transitions $\ket{i}_n \leftrightarrow \ket{e}_n$ and $\ket{e}_n \leftrightarrow \ket{g}_n$ to the laser and cavity fields with Rabi frequency $\Omega$ and vacuum Rabi frequency $g$, respectively. $ \hat S_{\alpha \beta} = \sum_n \hat \sigma^{(n)}_{\alpha \beta}/\sqrt{N}$ are collective operators that couple the internal states of each pair $n$ via $\hat \sigma_{\alpha,\beta}^{(n)} = \ket{\alpha}\bra{\beta}_n$  ($\alpha,\beta = i,e,g,x$). $\hat H$ is defined in a rotating frame \cite{SOM}\nocite{chase_collective_2008}\nocite{Bolanos_Algebraic_2015}\nocite{Hartmann_Generalized_2016}\nocite{gegg_efficient_2016}\nocite{Shammah_Open_2018}\nocite{Rom_State_2004}\nocite{Covey_Controlling_2017} with the detunings of the laser and the cavity, $\Delta=\omega_{ie}-\omega_L$ and $\delta = \omega_C -\omega_L - \omega_{gi}$, respectively. These are included in $\hat H_0 = \Delta \hat N_e + \delta \hat a^\dag \hat a$, where $\hat N_\alpha = \sum_n \hat \sigma_{\alpha \alpha}^{(n)}$ are total state populations, $\hat a$ is the cavity photon annihilation operator, and $\omega_L$, $\omega_C$ and
$\omega_{\alpha\beta}$ are the frequencies of the laser, the cavity and the transitions, respectively.

Dissipative terms are described by the super-operator
\begin{align}
	\mathcal{D} \hat{\rho} =  \mathcal{L}[\hat{L}_\kappa]\hat{\rho} + \sum_{n=1}^N \big( \mathcal{L}[\hat{L}^{(n)}_{\gamma_{i}}] + \mathcal{L}[\hat{L}^{(n)}_{\gamma_{x}}] + \mathcal{L}[\hat{L}^{(n)}_{\gamma_{g}}] \big) \hat{\rho}
	\label{eq:dissipator}
\end{align}
with $3N+1$ decay channels, each governed by a Lindblad term $\mathcal{L}[\hat L]\hat \rho = - \{\hat L^\dagger \hat L, \hat \rho\} + 2 \hat L \hat \rho \hat L^\dagger $. Here we include cavity decay with rate $2\kappa$, $ \hat{L}_\kappa = \sqrt{\kappa}\hat{a} $, and spontaneous emission from the excited state $\ket{e}_n$ 
for each pair $n$, $\hat{L}^{(n)}_{\gamma_{\alpha}} = \sqrt{\gamma_{\alpha}} \hat{\sigma}_{\alpha e}^{(n)}$ with rates $2\gamma_\alpha$ for $\alpha = i,g,x$. We define $\Gamma = \sum_\alpha \gamma_\alpha$ and the complex detunings $ \tilde{\Delta} = \Delta - \mi \Gamma $ and $ \tilde{\delta} = \delta - \mi \kappa$.
\medskip

In the regime of strong dissipation, both the excited states and the cavity mode  are weakly populated $\langle \hat N_{e} + \hat a^\dag \hat a \rangle \ll 1$ and can be adiabatically eliminated~\cite{reiter_effective_2012,SOM}. Then, the dynamics  reduces to an effective master equation for the sub-systems $\{ \ket{i}_n, \ket{g}_n, \ket{x}_n \}$ [see Fig.~\ref{fig:setup}(c)]. We find that the new effective Lindblad operators read
\begin{align}
	\hat L_\mathrm{eff}^\kappa 
	&= \sqrt{\lambda_\kappa} \hat{\xi} \hat{S}_{gi} \quad
	\hat L_\mathrm{eff}^{{\alpha}, {(n)}} 
	=  \sqrt{\lambda_\gamma^\alpha}  \left( \hat{\sigma}^{(n)}_{\alpha i} - \hat{\sigma}^{(n)}_{\alpha g} \hat{\xi} \hat{S}_{gi} \right) \label{eq:eff_lindblad_channels}
\end{align}
The terms $\hat L_\mathrm{eff}^\kappa$ and $\hat L_\mathrm{eff}^{{\alpha}, {(n)}}$ in Eq.~\eqref{eq:eff_lindblad_channels} result from a virtual excitation of the states $|e\rangle_n$ being lost via the cavity or via spontaneous emission, respectively. Here, $\lambda_\kappa = {\Omega^2 \kappa}/{g^2}$ and $\lambda_\gamma^\alpha = \Omega^2 \gamma_\alpha/ \tilde{\Delta}^2$ are the respective rates, while $\hat{\xi} = \sqrt{N} g^2 ( \hat{N}_{g} g^2 - \tilde{\Delta}\tilde{\delta} )^{-1}$ is a collective dimensionless operator stemming from the excited state propagator, which captures the effects of virtually excited superradiant states [in the weak light-matter coupling regime $(N_g+1)g^2 < (\kappa-\Gamma)^2/4$] or virtually excited polaritons [in the strong coupling regime $(N_g+1)g^2 > (\kappa-\Gamma)^2/4$]. Thus, Eq.~\eqref{eq:eff_lindblad_channels} gives rise to collective, dissipative, and uni-directional population transfer from the states $\ket{i}_n$ to the desired molecular bound states $\ket{g}_n$ and the loss states $\ket{x}_n$ [see Fig.~\ref{fig:setup}(c)], with rates that depend on the many-body state via $\hat \xi \hat S_{gi}$. 

We find a new effective Hamiltonian
\begin{align}
	\hat H_\mathrm{eff} &= -\frac{\Omega^2}{2\tilde{\Delta}} \left( \hat{N}_{i} + \sqrt{N} \hat{S}_{ig} \hat{\xi} \hat{S}_{gi}  \right) + \text{h.c.}  \label{eq:eff_ham} 
\end{align}
The first term $-\Omega^2 \hat N_i/(2\tilde \Delta)$ in Eq.~\eqref{eq:eff_ham} corresponds to the usual AC Stark shift for a small coupling $\Omega$. The second term corresponds to the self energy due to a molecule being virtually excited by the laser and exchanging this excitation with the cavity. Since  $[\hat N_\alpha$,  $\hat H_\mathrm{eff}]=0$, $\hat H_\mathrm{eff}$ cannot drive any coherent population transfer and thus we find that all interesting dynamics is driven by dissipation. In the following, we simulate the effective equations of motion first on bare resonance $\Delta = \delta = 0$, then on resonance with a (virtual) polariton. 
\medskip

Numerically, the master equation evolution with terms from Eqs.~\eqref{eq:eff_lindblad_channels} and \eqref{eq:eff_ham} can be efficiently simulated by exploiting the permutation symmetry among the $N$ three level systems, which allows for utilizing a collective spin basis \cite{zhang_monte-carlo_2018}. In practice we furthermore employ a quantum trajectory method~\cite{zhang_monte-carlo_2018,Daley_Quant_2014,SOM}. 
In the numerical simulations, the initial state is the product $\bigotimes_n \ket{i}_n$. 

For $\Delta = 0$, we choose typical parameters for RbCs as measured in Ref.~\cite{Debatin_Molecular_2011} [see also Fig.~\ref{fig:setup}(b) and Ref.~\cite{SOM}]. We consider up to $N=10^5$ molecules trapped in a three-dimensional optical lattice created by a laser with wavelength $\lambda_\mathrm{latt} = \SI{1064.5}{\nano\metre}$. Two lattice beams are placed at angles $\pm \theta$ ($\theta = \arccos[\lambda_\mathrm{latt}/(2\lambda_\mathrm{eg})] = \SI{57}{\degree}$) with respect to the cavity axis in order to match a desired cavity mode [Fig.~\ref{fig:setup}(a) and below].
The excited state has a half linewidth $\Gamma/2\pi = \SI{2.65}{\mega\hertz}$. The branching ratios $f_\alpha \equiv \gamma_\alpha/\Gamma$ for the decay from $\ket e$ into the states $\ket x$, $\ket g$, and $\ket i$ are $f_x \approx 0.999$,  $f_g = \num{1.3e-3}$, and  $f_i = \num{1.3e-4}$, respectively, such that photoassociation without a cavity leads to a maximal asymptotic value of $(\langle \hat N_g \rangle / N) (t\rightarrow \infty) \equiv (N_g / N) (t\rightarrow \infty) \equiv N_g^\infty / N = f_g / (f_g + f_x) \approx \num{1.3e-3}$. The  photoassociation laser (wavelength of $\lambda_\mathrm{PA} = \SI{1557}{\nano\metre}$) has a Rabi frequency $\Omega/2\pi = \SI{70}{\kilo\hertz}$ in the weak coupling regime. We assume a cavity of length $L = \SI{280}{\micro\metre}$, free spectral range $c/2L = \SI{535}{\giga\hertz}$, mode waist $\omega_0 = \SI{12}{\micro\metre}$, and half linewidth $\kappa/2\pi = \SI{5.4}{\mega\hertz}$, which is tuned in resonance with the $\lambda_\mathrm{eg} = \SI{977}{\nano\metre}$ transition $\ket e \leftrightarrow \ket g$, resulting in a peak vacuum Rabi frequency $g/2\pi = d_\mathrm{el}\sqrt{f_g\omega_{ge} / 2\hbar\varepsilon_0 V}/2\pi = \SI{770}{\kilo\hertz}$ with the mode volume $V = \pi \omega_0^2 L / 4$ and the electronic transition dipole moment $d_\mathrm{el}=0.1$\,a.u.~\cite{Debatin_Molecular_2011}. 
We assume the temperature to be small enough so that all molecules are in the lowest lattice band. For a typical lattice depth of $E_0 = 48E_R$~\cite{Takekoshi_Ultracold_2014}, with $E_R = (2\pi\hbar/\lambda_\mathrm{latt})^2/(2m_\mathrm{RbCs})$ the recoil energy, this implies $T \ll$\SI{400}{\nano\kelvin}, but even for higher temperatures the scheme may be beneficial~\cite{SOM}.

\medskip

\begin{figure}
    \centering
	\includegraphics[width=\columnwidth]{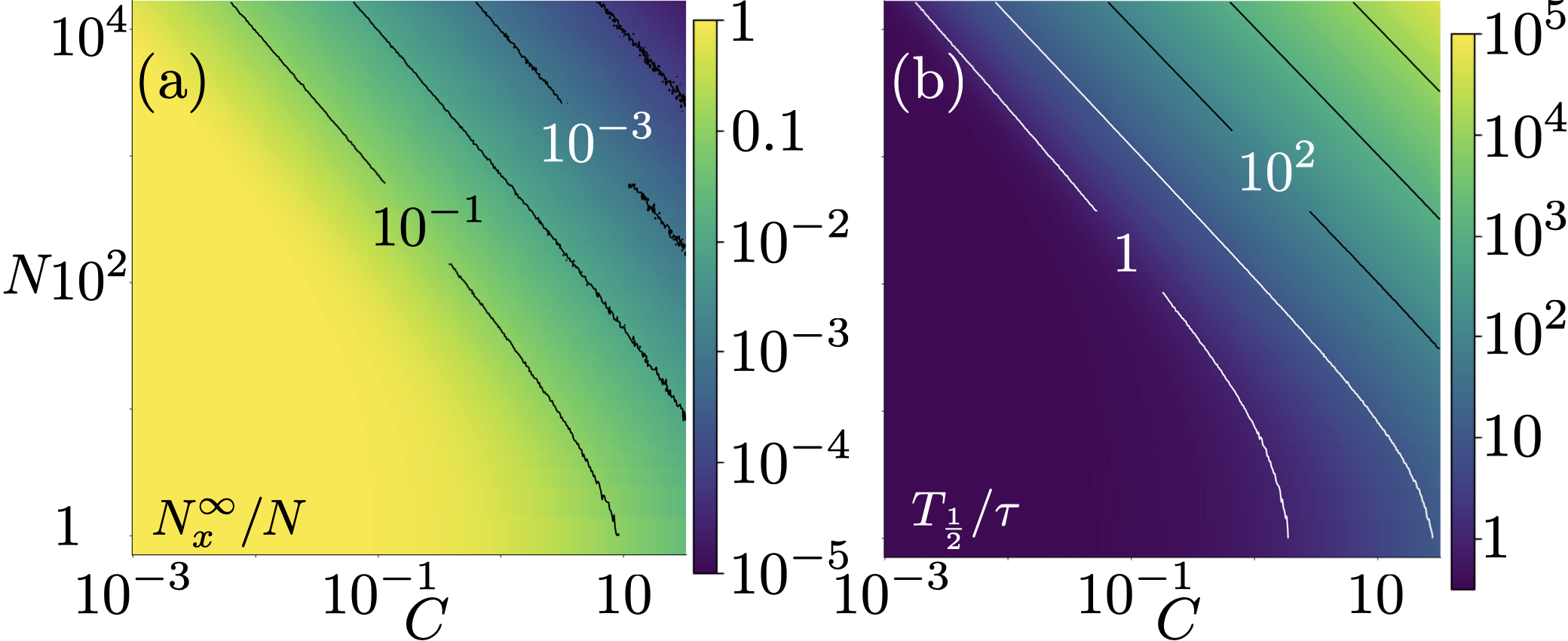}
	\caption{On bare resonance: (a) Contour plot of the final population fraction in the loss state $ \ket{x} $, $ N_x^\infty / N $, as a function of the number $ N $ of molecules in the cavity and the single molecule cooperativity $C$. (b) Contour plot of the time $T_\frac{1}{2}$ needed to transfer half of the population away from the state $ \ket{i} $ in $ \tau = \Gamma\Omega^{-2} $. All axes are logarithmic.}
	\label{fig:no_det}
\end{figure}

For $\Delta=\delta=0$, we find $\hat H_\mathrm{eff} = 0$
and the dynamics is governed by dissipative Lindblad terms {\em only}, with $\hat{\xi} = \sqrt{N} C (1 + \hat{N}_{g} C)^{-1}$. 
Figure~\ref{fig:setup}(d) shows exemplary results for the time evolution of the molecular ground state fraction $ N_g /N$ as a function of $N$, with $1\leq N \leq 10^5$ in units of the characteristic time scale $\tau = \Gamma\Omega^{-2}$.  For $N=1$ the figure shows that the presence of a cavity (here $C \approx 0.04$) induces an enhancement of $N_g^\infty/N$ from $\sim \SI{0.1}{\percent}$ (no cavity, dashed red line) to $\sim \SI{4}{\percent}$, due to increased state-selectivity~\cite{Kampschulte_Cavity_2018}.
Strikingly, with increasing $N$, we observe an enhancement towards $N_g^\infty/N \to 1$, at the cost of an increased transfer time. 
Figure~\ref{fig:no_det}(a) is a contour plot of the long-time population fraction $N_x^\infty/N$ in the loss state $\ket{x}$ as a function of $N$ and $C$. The plot shows that, for increasing collective cooperativity $NC$, $N_x^\infty/N$ rapidly decreases from its bare (no-cavity) value $\sim 1$ towards $0$ [upper right corner in Fig.~\ref{fig:no_det}(a)]. 

To gain further insight, we obtain an analytical solution of the dynamics in the limit of large collective cooperativity $NC \gg 1$ and large but finite molecule number $N_g \gg 1$. 
In the quantum trajectories picture, the decay rate of a state $\ket \psi$ is given by $\bra \psi -2\sum \hat L_\mathrm{eff}^\dagger \hat L_\mathrm{eff} \ket \psi$.
With these assumptions, we can restrict the discussion to the symmetric Dicke states, assume $(N_g+1)C \gg 1$, and neglect fluctuations by approximating operators by their expectation values $N_\alpha \equiv \langle \hat N_\alpha \rangle$. We then obtain the following rates for the decays via the different channels~\cite{SOM}
\begin{align}
    2\left\langle \hat L_\mathrm{eff}^{\kappa \dagger} \hat L_\mathrm{eff}^\kappa \right\rangle \approx \frac{2}{\tau} \frac{N_i}{(N_g + 1)C} \equiv \zeta_\kappa \label{eq:cav_dec} \,,
\end{align}
\begin{align}
    2 \sum_n \left\langle   \hat L_\mathrm{eff}^{\alpha, (n) \dagger} \hat L_\mathrm{eff}^{\alpha, (n)} \right\rangle  \approx \frac{2f_\alpha}{\tau} \frac{N_i}{(N_g+1)^2 C^2} \equiv \zeta_\alpha \,. \label{eq:spon_dec}
\end{align}
For $(N_g+1) C \gg 1$, the cavity-decay dominates, $\zeta_\kappa \gg \zeta_\alpha$. Dynamics is then governed by the non-linear rate equations $\dot N_i = -\zeta_x - \zeta_g - \zeta_{\kappa}$, $\dot N_x = \zeta_x$, and $\dot N_g = \zeta_g+\zeta_\kappa$, for which we provide analytical solutions in Ref.~\cite{SOM} for the time-dependence of the populations, $N_\alpha(t)$. 
For large $N_g^\infty$, we find for the loss state fraction
\begin{align}
    \frac{N_x^\infty}{N}\approx \frac{f_x \ln(N)}{NC} \xrightarrow{NC \rightarrow \infty} 0,
    \label{eq:xstate_scaling}
\end{align}
demonstrating a collective improvement over the single molecule result $f_x/(C+1)$ of Ref.~\cite{Kampschulte_Cavity_2018}.
The half time $T_\frac{1}{2} = \int_{N_i=N}^{N/2} \d N_i / \dot{N}_i$ for population transfer out of state $\ket{i}$ is well approximated by 
\begin{align}
    T_\frac{1}{2}\sim NC\tau. 
    \label{eq:halftime_bare}
\end{align}
This scaling is observed as straight contours for large $NC$ in the numerical simulations of Fig.~\ref{fig:no_det}(b). The demonstration of increased molecular yield in the ground state due to collective dissipative effects, at the cost of decreased transfer rates, is one of the central results of this work.

\medskip

\begin{figure}
	\centering
	\includegraphics[width=\columnwidth]{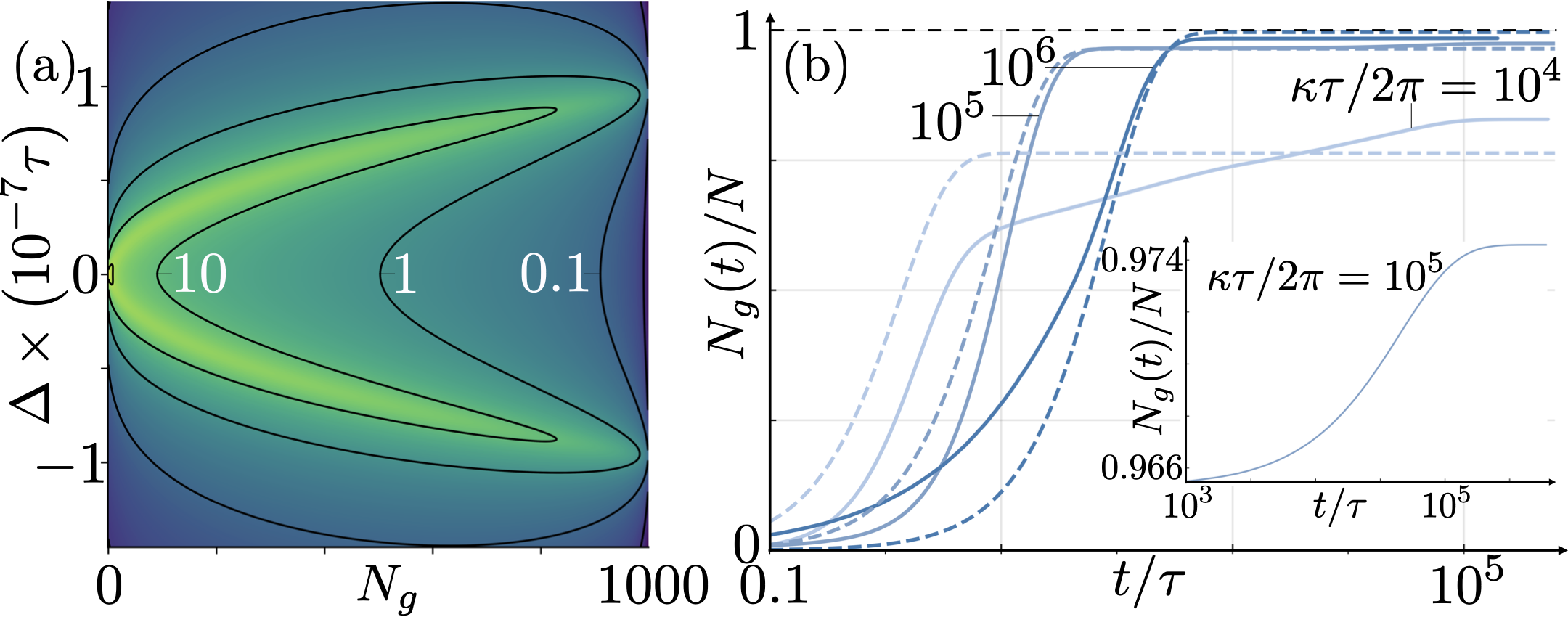}
	\caption{Chirped pulse: (a) Contour plot of the decay rate of Feshbach molecules $\dot{N}_i$ (in units $ \tau^{-1}$, for symmetric Dicke states), as a function of the laser detuning and the number of ground state molecules. The cavity is kept at resonance with the transition energy ($ \delta = \Delta $). (b) Simulated time evolution of the ground state population for different cavity decay rates $\kappa$. The parameters are chosen for $10^3$ Rb$_2$ molecules inside a cavity~\cite{Kampschulte_Cavity_2018, bellos_formation_2011}, i.\,e. $\Gamma/2\pi = \SI{6}{\mega\hertz}$ and $g/2\pi \approx \SI{50}{\mega\hertz}$. Dashed lines: analytical fits of Eqs.~\eqref{eq:pol_evolni} and \eqref{eq:pol_evolng}.}
	\label{fig:polaritons}
\end{figure}

We find that the slowdown of $T_\frac{1}{2}$ in Eq.~\eqref{eq:halftime_bare} is due to terms $\propto1/\qty[\qty(N_g+1) C]$ in Eqs.~\eqref{eq:cav_dec} and~\eqref{eq:spon_dec}, caused by Zeno blocking of the virtually excited superradiant states and detuning from the virtually excited polaritons. The latter usually dominates and is captured by Fig.~\ref{fig:polaritons}(a), which is a contour plot of $\dot N_i$ as a function of $N_g$ and $\Delta$, with $\delta = \Delta$. For $\Delta=0$, the figure shows that $\dot N_i$ decreases rapidly with increasing $N_g$. The rate $\dot N_i$ is instead maximized for an optimal choice of detuning
\begin{align}
    \Delta_\mathrm{opt}^{\pm} &= \pm \left[\max\qty(0, \qty(N_g+1)g^2 - \frac{\Gamma^2 + \kappa^2}{2})\right]^{1/2} \label{eq:polariton}.
\end{align}
This reflects the formation of two polaritons with energy $E^\pm \sim \Delta_\mathrm{opt}^{\pm}$ for large enough $N_g \geq (\Gamma^2 + \kappa^2) / 2g^2$. 
To circumvent the slowdown, we propose to chirp the laser detuning to stay resonant with the polariton energy, which depends on the (time dependent) ground state population $N_g(t)$. This adjustment can be adiabatic since the dynamics of $N_g(t)$ is  slow compared to $\Gamma$ [$\mathcal{O}\qty(\Omega^2 / \Gamma)$], and thus it is sufficient to consider a time dependent $\Delta(t)$ and $\delta(t) = \Delta(t)$ in Eqs.~\eqref{eq:eff_lindblad_channels} and \eqref{eq:eff_ham}.

For $g \ll \kappa + \Gamma \ll \sqrt{N_g+1}g$, the decay rates of the different channels assume the simple form~\cite{SOM}{\footnote{We note that the inequality $g \ll \kappa + \Gamma$ ensures that staying in resonance with the polariton is possible, as for $g \gtrsim \kappa + \Gamma$ this is prevented by fluctuations of $N_g$. For small $N_g$ we have $\kappa + \Gamma \gtrsim \sqrt{N_g+1}g$ and no polariton splitting is present.}}
\begin{align}
    2\left\langle \hat L_\mathrm{eff}^{\kappa \dagger} \hat L_\mathrm{eff}^\kappa \right\rangle &\approx \frac{2\Omega^2 \kappa}{\qty(\kappa + \Gamma)^2} N_i \equiv \zeta_\kappa \label{eq:pol_cav_dec} \\
    2 \sum_n \left\langle \hat L_\mathrm{eff}^{\alpha, (n) \dagger} \hat L_\mathrm{eff}^{\alpha, (n)} \right\rangle &\approx \frac{2\Omega^2 \gamma_\alpha}{\qty(\kappa + \Gamma)^2} N_i \equiv \zeta_\alpha.
    \label{eq:pol_spon_dec}
\end{align}
and the rate equations $\dot N_{i,g,x}$ above are solved as
\begin{align}
    \frac{N_i(t)}{N} &= \exp[-\frac{2\Omega^2 \qty(\kappa + \gamma_g + \gamma_x)t}{\qty(\kappa + \Gamma)^2}] \label{eq:pol_evolni} \\
    \frac{N_g(t)}{N} &= \frac{\kappa + \gamma_g}{\kappa + \gamma_g + \gamma_x} \qty{1 - \exp[-\frac{2\Omega^2 \qty(\kappa + \gamma_g + \gamma_x)t}{\qty(\kappa + \Gamma)^2}]}. \label{eq:pol_evolng}
\end{align}
These results are formally similar to those of Ref.~\cite{Perez_Ultracold_2017} for a single molecule coupled to a photon wave-guide, however, here the cavity decay rate $\kappa$ is fully tuneable.
We note that while collective effects are present in the polariton formation, the final rate is  independent of $N$ {\footnote{This is in contrast to the scheme of Ref.~\cite{Perez_Ultracold_2017}, for which collective effects decrease the efficiency due to dark state population caused by individual laser excitations.}.}
For $\kappa \gg \gamma_x$, the ground state population approaches $N$ as  $N_g^\infty/N \simeq 1 - \gamma_x/\kappa$, at the cost of an increasing time-scale $\sim \kappa/\Omega^2$, due to the continuous Zeno effect~\cite{Gardiner_2015_Quantum}. 

Figure~\ref{fig:polaritons}(b) shows a comparison of numerical and analytical results (continuous and dashed lines, respectively) for the increase of $N_g/N$ as a function of time $t$, for different values of $\kappa$. We find good agreement for large values of $t/\tau$ and $\kappa \tau$ in the regime of validity of Eqs.~\eqref{eq:pol_evolni} and \eqref{eq:pol_evolng}, as expected. In addition to this dynamics, a few molecules are trapped in so-called ``dark states'' $\ket d$ that cannot decay via the cavity ($\hat L_\mathrm{eff}^\kappa \ket d = 0$). This minor effect is caused by the breaking of permutation symmetry through spontaneous emission~\cite{SOM,Shammah_Superradiance_2017,Gegg_2018_Superradiant}, and responsible for a time delay in reaching the asymptotic $N_g^{\infty}$, as magnified in the inset of Fig.~\ref{fig:polaritons}(b).

\smallskip

Whether higher molecular yields are reached by staying on bare resonance or chirping the laser depends on what limits state selectivity. If transfer times are not a concern, staying on bare resonance is usually best, as we estimate $N_x^{\infty}({\rm chirp})> N_x^{\infty}({\rm bare})$ for $Ng^2/[\kappa^2 \ln(N)]> 1$,~\cite{SOM}. If instead transfer times are a concern, due, e.g., to background gas collisions, then the chirped scheme may be better, as for a given $T_{\frac{1}{2}}$ we find $N_x^{\infty}({\rm bare}) \sim \ln(N) N_x^{\infty}({\rm chirp})$~\cite{SOM}. These behaviors, derived for identical cavity coupling strengths $g_n$, hold also approximately for moderately varying $g_n$ due to, e.g, in-homogeneity of the cavity mode, lattice geometry or thermal motion~\cite{SOM}.

For $N=10^4$ RbCs Feshbach molecules (see parameters above), the system is closer to the first scenario and we find that staying on bare resonance ($\Delta=0$) provides the highest yield. For a reasonable lattice lifetime of \SI{1}{\second}, we obtain a peak ground state population $N_g / N \approx \SI{92}{\percent}$ after \SI{55}{\milli\second} with a transfer half time $T_\frac{1}{2} \approx \SI{3.2}{\milli\second}$ (\SI{98}{\percent} for infinite lattice lifetime). These results are essentially unchanged by considering locally different coupling constants $g_n = g(\vec x_n) = g \exp[-(x_n^2+y_n^2)/\omega_0^2]\cos(2\pi z_n /\lambda_\mathrm{eg})$, due to the finite cavity mode waist $\omega_0$ and the different lattice positions, with $z_n$ ($x_n$, $y_n$) oriented along the cavity axis (in the perpendicular planes) [see Fig.~\ref{fig:setup}(a)]. For example, for a $20 \times 20 \times 25$ lattice at angle $\theta = \SI{57}{\degree}$ and assuming perfect matching of lattice and cavity modes with $\cos(2\pi z_n /\lambda_\mathrm{eg}) = 1$
\footnote{Lattice positions are defined as $x_n = (m+1/2)\lambda_\mathrm{latt} / 2$,  and $y_n = (m'+1/2)\Delta y$ with $\Delta y = \lambda_\mathrm{latt} / [4\sin(\theta)] \approx \SI{317}{\nano\metre}$, with $-10 \leq (m,m') < 10$ integers.}, 
we find a peak $N_g / N \sim \SI{92}{\percent}$, a transfer time \SI{48}{\milli\second} and $T_\frac{1}{2}\sim\SI{2.7}{\milli\second}$, with infinite lattice lifetime final fraction \SI{97}{\percent}.
Thus, ground state populations comparable to STIRAP ($\sim \SI{90}{\percent}$)~\cite{Lang_Ultracold_2008, Takekoshi_Ultracold_2014} can be achieved without the need of time-dependent laser pulses. These results are  robust against reasonable lattice mismatches. Even in a worst case scenario of complete positional disorder [i.e., uniform and Gaussian ($\sigma_{xy} = \SI{5}{\micro\metre}$) distributions in the $z$ and $x-y$ directions, respectively] 
we find a peak $N_g / N \sim \SI{71}{\percent}$ (\SI{73}{\percent} for infinite lattice lifetime) after \SI{21}{\milli\second} and $T_\frac{1}{2}\sim\SI{0.4}{\milli\second}$. 

For a scenario with $10^3$ Rb$_2$ Feshbach molecules (see Fig.~\ref{fig:polaritons}, parameters as in Ref.~\cite{bellos_formation_2011, Kampschulte_Cavity_2018}) we are in the regime where the chirped pulse results in a higher yield. For example, choosing a vacuum Rabi frequency $g/2\pi \approx \SI{50}{\mega\hertz}$, a laser Rabi frequency $\Omega/2\pi = \SI{200}{\kilo\hertz}$, and a cavity half linewidth $\kappa/2\pi = \SI{300}{\mega\hertz}$, we obtain a ground state population of $N_g / N \approx \SI{98}{\percent}$ after \SI{5}{\milli\second} ($T_\frac{1}{2}\sim \SI{0.5}{\milli\second}$). Even for a spatially disordered worst case scenario (uniform position distribution along $z$, $\sigma_{xy}$ = \SI{2.5}{\micro\metre}, mode waist of \SI{5}{\micro\metre}), we reach a ground state fraction of \SI{89}{\percent} after \SI{5}{\milli\second} ($T_\frac{1}{2}\sim \SI{0.3}{\milli\second}$). In both cases this is a significant increase from \SI{54}{\percent} without cavity, and can overcome typical STIRAP efficiencies~\cite{Lang_Ultracold_2008}.

Similar to STIRAP~\cite{Takekoshi_Ultracold_2014,vitanov2017stimulated}, the presence of additional excited states in proximity to the $\ket e$ state might decrease the transfer efficiency of our schemes. This can be avoided by choosing an excited state with sufficiently large hyperfine and Zeeman splitting~\cite{Takekoshi_Ultracold_2014}. Once the rovibrational ground state is reached, population transfer between the hyperfine sublevels can be achieved with high fidelity~\cite{Ospelkaus2010Jan}.

\medskip

In summary, we proposed two novel methods for high-yield state selective preparation of ultracold molecules in a cavity that exploit collective and dissipative effects.
It is an exciting prospect to investigate how similar collective effects could be used to engineer generic state-transfer schemes and even chemical reactions outside of the ultracold regime~\cite{Hutchison_Modify_2012,Thomas_Tiltin_2019,Herrera2020Mar,Torma_Strong_2014,Ebbesen_Hybrid_2016}, such as room-temperature cavity-modified electron transfer reactions~\cite{Mandal2020Polariton,Wellnitz2020disschem}. The experimental setups proposed here -- molecules trapped on a lattice potential and embedded in a cavity --  offer unique opportunities to explore collective dynamics for measurements~\cite{Niezgoda2020Cooperatively}, non-equilibrium quantum phase transitions~\cite{Barberena2019Driven,Muniz2020Exploring}, or quantum information applications using long-lived molecular states~\cite{rabl2006hybrid} and cavity-controlled gates~\cite{Imamoglu1999quantum,Majer2007Coupling}, while also allowing for a non-destructive detection of the molecules~\cite{sawant2018detection,zhu2020resonator}.

\medskip
\medskip

We thank Claudiu Genes for helpful discussions. This work is supported by ANR 5 ``ERA-NET QuantERA'' - Projet
``RouTe'' (ANR-18-QUAN-0005-01), LabEx (``Nanostructures in Interaction with their Environment,'' NIE) under contract ANR-11-LABX-0058 NIE within the Investissement d’Avenir program ANR-10-IDEX-0002-02, and IdEx Unistra project STEMQuS. D.~W. acknowledges financial support from Agence Nationale de la Recherche (Grant ANR-17-EURE-0024 EUR QMat).  G.~P. acknowledges support from the Institut Universitaire de France (IUF) and the University of Strasbourg Institute of Advanced Studies (USIAS).

\bibliographystyle{apsrev4-2}
\bibliography{colldissform}

\appendix

\begin{widetext}

\newpage

\begin{center}
\textbf{\large Supplemental Materials: Collective Dissipative Molecule Formation in a Cavity}
\end{center}

\setcounter{equation}{0}
\setcounter{figure}{0}
\setcounter{table}{0}
\setcounter{page}{1}
\setcounter{secnumdepth}{2}
\makeatletter
\renewcommand{\theequation}{S\arabic{equation}}
\renewcommand{\thefigure}{S\arabic{figure}}

In Sec.~\ref{sec:app_rotating} we define the rotating frame. In Sec.~\ref{sec:app_elim}, we give details for the adiabatic elimination. In Sec.~\ref{sec:app_numerics}, we discuss how the permutation symmetry can be used for an efficient simulation and display the resulting equations that are used for the simulation. In Sec.~\ref{sec:app_states} we give the states used for the RbCs simulations. In Sec.~\ref{sec:app_rate_eqs} we derive the rate equations on bare resonance, and their solution is presented in Sec.~\ref{sec:app_rate_solution}. The rate equations on polariton resonance and the function to match the polariton resonance are provided in Sec.~\ref{sec:app_polaritons}. In Sec.~\ref{sec:app_dark} we give some details on the dark states. In Sec.~\ref{sec:app_comp} we compare the schemes on bare resonance and on polariton resonance, and in Sec.~\ref{sec:app_disorder} we display the threshold and effective cooperativity used to describe the local cavity coupling constants. In Sec.~\ref{sec:app_finT} we discuss the effects of finite temperature.

\section{Rotating Frame} \label{sec:app_rotating}

Setting the initial state energy to zero, the full Hamiltonian reads in a non-rotating frame of reference
\begin{align}
    \hat H = \sum_n \omega_{ie}\hat \sigma^{(n)}_{ee} + \omega_{ix}\hat \sigma^{(n)}_{xx} - \omega_{gi} \hat \sigma^{(n)}_{gg} + \omega_C \hat a^\dagger \hat a + \Omega \{\hat \sigma^{(n)}_{ie}\exp[\mi(\omega_Lt-\vec k\vec x_n)] + \mathrm{h.c.}\} + g (\hat \sigma^{(n)}_{ge}\hat a^\dagger + \mathrm{h.c.})
\end{align}
We arrive at the given Hamiltonian by applying the unitary transform
\begin{align}
    \mathcal U = \exp[\mi\qty(\sum_n \omega_Lt\hat \sigma^{(n)}_{ee} + \omega_{ix}t\hat \sigma^{(n)}_{xx} - \omega_{gi}t\hat \sigma^{(n)}_{gg} + (\omega_L + \omega_{gi})t\hat a^\dagger \hat a + \vec k \vec x_n \hat \sigma^{(n)}_{ee})]
\end{align}
according to the rules
\begin{align}
    \hat H \rightarrow \hat H' &= \mathcal U \hat H \mathcal U^\dagger + \mi \qty(\partial_t \mathcal U) \mathcal U^\dagger \\
    \ket \psi \rightarrow \ket{\psi'} &= \mathcal U \ket \psi \\
    \hat O \rightarrow \hat O' &= \mathcal U \hat O \mathcal U^\dagger
\end{align}
with $\hat O$ an arbitrary observable. In the paper the further definitions $\Delta = \omega_{ie}-\omega_L$ and $\delta = \omega_C - \omega_L - \omega_{gi}$ were used.

\section{Adiabatic Elimination}
\label{sec:app_elim}

In order to adiabatically eliminate the cavity and the excited states, we follow the formalism of \textit{Reiter and S{\o}rensen} \cite{reiter_effective_2012}. First, we split the system into an excited state manifold with fast dynamics and a ground state manifold with slow dynamics, both of which are weakly coupled. We define the ground state manifold by $\hat N_e + \hat a^\dagger \hat a = 0$, so that there are neither molecular excitations nor photons. The excited state manifold contains all remaining states ($\hat N_e + \hat a^\dagger \hat a \geq 1$), but as we use the interaction as a perturbation, it is sufficient to restrict the analysis to the single excitation limit $\hat N_e + \hat a^\dagger \hat a = 1$. The condition for the adiabatic elimination to be valid is that the interaction is much slower than the excited state dynamics, e.\,g. because $\sqrt{N}\Omega \ll \Gamma$ or $\sqrt{N}\Omega \ll \sqrt{N_g}g$. Following the notation of \textit{Reiter and S{\o}rensen}, we arrive at:
\begin{align}
    \hat H_e &= \hat H_0 + \hat H_C \nonumber \\
    &= \Delta \hat N_e + \delta \hat a^\dagger \hat a + g \sqrt{N} \qty(\hat a^\dagger \hat S_{ge} + \hat S_{eg} \hat a) \\
    \hat H_g &= 0 \\
    \hat V^+ &= \sqrt{N}\Omega \hat S_{ei} \\
    \hat V^- &= \sqrt{N}\Omega \hat S_{ie} 
\end{align}
The Lindblad operators $\hat L_k$ are defined in the main paper.

Next, we calculate the non-hermitian Hamiltonian $\hat H_\mathrm{NH} \equiv \hat H_e - \mi \sum \hat L^\dagger_k \hat L_k$. Note that the factor 2 compared to \textit{Reiter and S{\o}rensen} \cite{reiter_effective_2012} arises due to a different definition of the Lindblad operators:
\begin{align}
    \hat H_\mathrm{NH} &= \tilde \Delta \hat N_e + \tilde \delta \hat a^\dagger \hat a + g \sqrt{N} \qty(\hat a^\dagger \hat S_{ge} + \hat S_{eg} \hat a)
\end{align}
with $\tilde \Delta \equiv \Delta - \mi \Gamma$ and $\tilde \delta \equiv \delta - \mi \kappa$. In the single excitation limit $\hat H_\mathrm{NH}$ can be inverted. It is straightforward to confirm that:
\begin{align}
    \hat H_\mathrm{NH}^{-1} &= \qty[\tilde{\Delta}^2\tilde{\delta} - \tilde{\Delta}(\hat N_g + \hat{N}_e)g^2]^{-1} \left\{\tilde{\Delta}^2\hat{a}^\dagger\hat{a} - g\sqrt{N}\tilde{\Delta} \qty(\hat S_{eg}\hat{a} + \hat S_{ge} \hat{a}^\dagger) + \qty[\tilde{\Delta}\tilde{\delta} - \qty(\hat{N}_g + 1) g^2 ] \hat N_e + Ng^2 \hat S_{eg}\hat S_{ge}\right\}
\end{align}

The effective operators are now given by:
\begin{align}
	\hat{H}_\mathrm{eff} &= - \frac{1}{2} \hat{V}^- \left[ \hat{H}_\mathrm{NH}^{-1} + \left( \hat{H}_\mathrm{NH}^{-1} \right)^\dagger \right] \hat{V}^+ + \hat H_g \\
	\hat{L}_\mathrm{eff}^k &= \hat{L}_k \hat{H}_\mathrm{NH}^{-1} \hat{V}^+
\end{align}
We find:
\begin{align}
	\hat H_\mathrm{eff} &= -\frac{\Omega^2}{2\tilde{\Delta}} \left( \hat{N}_{i} + \sqrt{N} \hat{S}_{ig} \hat{\xi} \hat{S}_{gi}  \right) + \text{h.c.} \\
	\hat L_\mathrm{eff}^\kappa 
	&= \frac{\Omega\sqrt{\kappa}}{g} \hat{\xi} \hat{S}_{gi} \\
	\hat L_\mathrm{eff}^{{\alpha}, {(n)}} 
	&=  \frac{\Omega\sqrt{\gamma_\alpha}}{\tilde \Delta}  \left( \hat{\sigma}^{(n)}_{\alpha i} - \hat{\sigma}^{(n)}_{\alpha g} \hat{\xi} \hat{S}_{gi} \right)
\end{align}
with $\hat{\xi} = \sqrt{N} g^2 \qty( \hat{N}_{g} g^2 - \tilde{\Delta}\tilde{\delta} )^{-1}$.

The adiabatic elimination as discussed above is valid in the single excitation limit, which can be assumed if $\langle \hat N_e + \hat a^\dagger \hat a \rangle \ll 1$. The number of excitation $\langle \hat N_e + \hat a^\dagger \hat a \rangle$ can be estimated by comparing the pumping rate $\sqrt{N_i}\Omega$ to the total excitation decay $\Gamma_\mathrm{tot}$ and the total detuning $\Delta_\mathrm{tot}$:
\begin{align}
    \langle \hat N_e + \hat a^\dagger \hat a \rangle \approx \frac{N_i\Omega^2}{\Gamma_\mathrm{tot}^2 + \Delta_\mathrm{tot}^2} \overset{!}{\ll} 1\,. \label{eq:elim-cond}
\end{align}

We first consider the scheme on bare resonance. The short times dynamics is best described in terms of bare excitons so that $N_i\approx N$, $\Gamma_\mathrm{tot} = \Gamma$, and $\Delta_\mathrm{tot} = 0$. In this case we can rewrite Eq.~\eqref{eq:elim-cond} to find $N \overset{!}{\ll} \Gamma^2/\Omega^2 \approx 1400$ for the RbCs parameters. In contrast, the long time dynamics is best described by polaritons and dark states. As the latter are populated only very slowly, we restrict the analysis of the long time dynamics to polaritons. We find $\Gamma_\mathrm{tot} = (\kappa + \Gamma) / 2$ and $\Delta_\mathrm{tot}^2 = N_gg^2 - (\Gamma-\kappa)^2/2$. For $g \gg \Gamma + \kappa$ and $N_g > 0$ the condition Eq.~\eqref{eq:elim-cond} simplifies to $N_g/N \ll g^2/\Omega^2 \approx \num{8e-3}$. We conclude for the RbCs parameters in the paper, although the initial dynamics is not described correctly by adiabatic elimination, after a population fraction of around \SI{1}{\percent} is transferred to the ground state, the adiabatic elimination is valid.

Using the chirped pulse the initial dynamics is still described by bare excitons as above and we find $N \overset{!}{\ll} \Gamma^2/\Omega^2 \approx 1000$ for the Rb$_2$ parameters. For later times, again ignoring dark states, we find $\Gamma_\mathrm{tot} = (\kappa + \Gamma) / 2$ and $\Delta_\mathrm{tot}^2 = 0$. This leads to the condition $(\kappa^2/2 + \Gamma^2/2) / \Omega^2 \approx \num{1.4e5} \overset{!}{\gg} N$. Thus for the Rb$_2$, the initial dynamics is not fully captured by the adiabatic elimination, but as soon as polaritons form, which happens after a short time, the elimination condition is clearly fulfilled.

Note that for the theoretical part of the paper the results are fully independent of the choice of $\Omega$, which only enters into the definition of $\tau$. Hence, for the theory the adiabatic elimination condition is fulfilled if $\Omega$ is kept small enough to fulfill condition Eq.~\eqref{eq:elim-cond}.

\section{\label{sec:app_numerics} Numerical Simulations}

In order to develop an efficient algorithm, we use two steps: In a first step we go from $N$ three level molecules (with states $\ket i$, $\ket g$, and $\ket x$) to two level molecules (with states $\ket i$ and $\ket g$) with variable molecule number. In a second step we take advantage of the permutation symmetry of the system to reduce $N$ spin-1/2 (two level) systems (dimension $\sim 2^N$) to one spin-$N/2$ system (dimension $\sim N^2$) \cite{chase_collective_2008,Bolanos_Algebraic_2015,Hartmann_Generalized_2016,gegg_efficient_2016,zhang_monte-carlo_2018,Shammah_Open_2018}. Note that both steps are exact.

For the first step we use that molecules that enter state $\ket x$ have no coherence with the rest of the system (decay only via local dissipation) and have no influence on the dynamics of the system. This allows us to treat decay into state $\ket x$ as molecule loss and we only need to treat the dynamics of the remaining two level systems.

In order to simulate $N$ identical two level systems with particle loss, we employ a quantum trajectories algorithm analogous to the one used by \textit{Zhang et al.} \cite{zhang_monte-carlo_2018}. We describe the dynamics of the two level systems in the Dicke basis $\ket{J, M}$ \cite{chase_collective_2008}, while keeping track of the molecule number $N$ \cite{zhang_monte-carlo_2018}. The result are equations for a matrix of the form $\overline{\ket{N,J,M}\bra{N,J,M'}}$.
The contributions to the equations of motion for the diagonal matrix elements $\overline{\ket{N,J,M}\bra{N,J,M}}$ are given in Tab.~\ref{tab:dicke_master}.

\begin{table}
    \renewcommand{\arraystretch}{1.8}
    \centering
    \begin{tabular}{|c|l|}
        \hline
        Term & Value \\
        \hline
        $-\mi \qty[\hat H_\mathrm{eff}, \overline{\ket{N,J,M}\bra{N,J,M}}]$ & 0 \\ 
        \hline
        \multirow{3}{5.2cm}{$-\qty{\sum_k \hat L_\mathrm{eff}^{k \dagger} L_\mathrm{eff}^k, \overline{\ket{N,J,M}\bra{N,J,M}}}$} 
        & $-\frac{2\Omega^2\kappa}{g^2} \overline{\ket{N,J,M}\bra{N,J,M}} (J+M)(J-M+1) \abs{\xi}^2 $ \\
        & $-\frac{2\Omega^2\Gamma}{\abs{\tilde \Delta}^2} \overline{\ket{N,J,M}\bra{N,J,M}} $ \\
        & $\qquad \times \qty[\frac{ N}{ 2} + M - 2(J+M)(J-M+1)\Re\qty(\xi) + (J+M)(J-M+1)\qty(\frac{ N}{ 2} - M + 1)\abs{\xi}^2]$ \\
        \hline
        $\hat L_\mathrm{eff}^\kappa \overline{\ket{N,J,M}\bra{N,J,M}} \hat L_\mathrm{eff}^{\kappa \dagger}$
        & \hspace{4pt} $\frac{ \Omega^2\kappa}{ g^2} \overline{\ket{N,J,M-1}\bra{N,J,M-1}} (J+M)(J-M+1) \abs{\xi}^2 $ \\
        \hline
        \multirow{4}{5.2cm}{$\sum_n \hat L_\mathrm{eff}^{i, (n)} \overline{\ket{N,J,M}\bra{N,J,M}} \hat L_\mathrm{eff}^{i, (n) \dagger}$} 
        & \hspace{4pt} $\frac{ \Omega^2\gamma_i}{ \abs{\tilde \Delta}^2}
        \overline{\ket{N,J-1,M}\bra{N,J-1,M}}
        \beta^J_N (J-M) $ \\
        &$+ \frac{ \Omega^2\gamma_i}{ \abs{\tilde \Delta}^2}
        \overline{\ket{N,J,M}\bra{N,J,M}} 
        \qty[ \frac{ N}{ 4} + M - (J+M)(J-M+1)\Re\qty(\xi) ]$ \\
        &$+ \frac{ \Omega^2\gamma_i}{ \abs{\tilde \Delta}^2}
        \overline{\ket{N,J,M}\bra{N,J,M}} \alpha_N^J \abs{M - (J+M)(J-M+1)\xi}^2 $ \\
        &$+ \frac{ \Omega^2\gamma_i}{ \abs{\tilde \Delta}^2}
        \overline{\ket{N,J+1,M}\bra{N,J+1,M}} \delta_N^J (J+M+1) $ \\
        \hline
        \multirow{4}{5.2cm}{$\sum_n \hat L_\mathrm{eff}^{g, (n)} \overline{\ket{N,J,M}\bra{N,J,M}} \hat L_\mathrm{eff}^{g, (n) \dagger}$} 
        & \hspace{4pt} $ \frac{ \Omega^2\gamma_g}{ \abs{\tilde \Delta}^2}
        \overline{\ket{N,J-1,M-1}\bra{N,J-1,M-1}}
        \beta^J_N (J+M-1) $ \\
        &$+ \frac{ \Omega^2\gamma_g}{ \abs{\tilde \Delta}^2}
        \overline{\ket{N,J,M-1}\bra{N,J,M-1}} (J+M)(J-M+1) \qty[ - \Re\qty(\xi) + \frac{ 1}{ 4}\abs{\xi}^2 ]$ \\
        &$+ \frac{ \Omega^2\gamma_g}{ \abs{\tilde \Delta}^2}
        \overline{\ket{N,J,M-1}\bra{N,J,M-1}}
        \alpha^J_N (J+M)(J-M+1)\abs{1 + (M-1)\xi}^2 $ \\
        &$+ \frac{ \Omega^2\gamma_g}{ \abs{\tilde \Delta}^2}
        \overline{\ket{N,J+1,M-1}\bra{N,J+1,M-1}}
        \delta^J_N (J-M+2) $ \\
        \hline
        \multirow{2}{5.2cm}{$\sum_n \hat L_\mathrm{eff}^{x, (n)} \overline{\ket{N,J,M}\bra{N,J,M}} \hat L_\mathrm{eff}^{x, (n) \dagger}$} 
        & \hspace{4pt} $\frac{ \Omega^2\gamma_x}{ \abs{\tilde \Delta}^2}
        \overline{\ket{N-1,J-\frac{ 1}{ 2},M-\frac{ 1}{ 2}}\bra{N-1,J-\frac{ 1}{ 2},M-\frac{ 1}{ 2}}} 4J\beta^J_N$\\
        & $+\frac{ \Omega^2\gamma_x}{ \abs{\tilde \Delta}^2}
        \overline{\ket{N-1,J+\frac{ 1}{ 2},M-\frac{ 1}{ 2}}\bra{N-1,J+\frac{ 1}{ 2},M-\frac{ 1}{ 2}}} 4(J+1)\delta^J_N$ \\
        \hline
    \end{tabular}
    
    \caption{Different contributions to the time evolution of density matrix element $\overline{\ket{N,J,M}\bra{N,J,M}}$. Notation: $\xi = g^2 / [(N/2 - M + 1)g^2 - \tilde \Delta \tilde \delta]$ (a factor of $\sqrt{N}$ different from the operator $\hat \xi$). $\alpha^J_N = (N+2) / (4J(J+1))$, $\beta^J_N = (N + 2J + 2)(J+M)\abs{1 - (J-M+1)\xi}^2 / (4J(2J+1))$, $\delta^J_N = (N-2J)(J-M+1)\abs{1+(J+M)\xi}^2 / [4(J+1)(2J+1)]$.}
    \label{tab:dicke_master}
\end{table}

As there are no off-diagonal elements generated in the equations in Tab.~\ref{tab:dicke_master} and the initial state is given by the diagonal element $\overline{\ket{N,N/2,N/2}\bra{N,N/2,N/2}}$, the diagonal elements are sufficient to describe the system dynamics. This motivates the description using a quantum trajectory algorithm, which becomes a simple Monte Carlo Markov chain of jumps between the matrix elements, due to the trivial Hamiltonian contribution.

\section{\label{sec:app_states} Molecular States}

We choose the states proposed in Ref.~\cite{Debatin_Molecular_2011} for STIRAP association of ultracold molecules. The initial state $\ket i$ is given by a Feshbach molecule in the sixth vibrational level below dissociation threshold, characterized by the atomic quantum numbers $f_\mathrm{Rb} = 2$; $m_{f_\mathrm{Rb}}=2$; $f_\mathrm{Cs} = 4$; $m_{f_\mathrm{Cs}}=2$; $m_f = 4$, and the quantum numbers for the atomic motion $l=2$; $m_l=0$, which couple to $M_F=4$. For the excited state we choose $\ket{e} = (A^1\Sigma^+ - b^3\Pi)0^+\ (v'=38)$, which has favorable transition dipole moments to both the initial Feshbach molecule and to the absolute ground state $\ket g = X^1\Sigma^+\, (v=0)$.

\section{\label{sec:app_rate_eqs} Derivation of Rate Equations}

In order to consider the effects of the different decay channels, we need to separate the decay rates given in the second row of Tab.~\ref{tab:dicke_master} into the different decay channels. For no detuning $\Delta = \delta = 0$, this yields:
\begin{align}
    -\qty{\hat L_\mathrm{eff}^{\kappa \dagger} L_\mathrm{eff}^\kappa, \overline{\ket{N,J,M}\bra{N,J,M}}} &= -\frac{2\Omega^2}{C\Gamma} \frac{(J+M)(J-M+1)}{\qty(\frac{N}{2} - M + 1 + \frac{1}{C})^2} \overline{\ket{N,J,M}\bra{N,J,M}} \\
    -\qty{\sum_n \hat L_\mathrm{eff}^{\alpha, (n) \dagger} L_\mathrm{eff}^{\alpha, (n)}, \overline{\ket{N,J,M}\bra{N,J,M}}} &= -\frac{2\Omega^2\gamma_\alpha}{\Gamma^2} \overline{\ket{N,J,M}\bra{N,J,M}} \times \nonumber \\
    &\mkern-72mu \left[\frac{N}{2} + M - 2\frac{(J+M)(J-M+1)}{\frac{N}{2} - M + 1 + \frac{1}{C}} + \frac{(J+M)(J-M+1)\qty(\frac{N}{2} - M + 1)}{\qty(\frac{N}{2} - M + 1 + \frac{1}{C})^2}\right]
\end{align}
where $\alpha = i, g, x$.

To simplify these equations, we make two assumptions: (i) We assume that the dynamics is taking place in the completely symmetric state, for which $J = N/2$, and (ii) we assume $N_gC \gg 1$. The first assumption is justified further below. The second assumption is justified for large collective cooperativity $NC$ and not too small $C$ (e.\,g. $C \sim 10^{-3}$ is fine for $N \sim 10^5$, but not thermodynamic limit with $N \rightarrow \infty$ and $C\rightarrow 0$). In this way, only the initial dynamics with $N_g \ll N$ is ignored, which we empirically find to be a good approximation. Using $N_g = N/2 - M$ and $N_i = N/2 + M$, we simplify:
\begin{align}
    -\qty{\hat L_\mathrm{eff}^{\kappa \dagger} \hat L_\mathrm{eff}^\kappa, \overline{\ket{N,J,M}\bra{N,J,M}}} &\overset{(i)}{\approx} -\frac{2\Omega^2}{C\Gamma} \frac{N_i \qty(N_g + 1)}{\qty(N_g + 1 + \frac{1}{C})^2} \nonumber \\
    &\overset{(ii)}{\approx} -\frac{2\Omega^2}{\Gamma} \frac{N_i}{\qty(N_g + 1)C} \label{eq:app_rates1} \\
    -\qty{\sum_n \hat L_\mathrm{eff}^{\alpha, (n) \dagger} \hat L_\mathrm{eff}^{\alpha, (n)}, \overline{\ket{N,J,M}\bra{N,J,M}}} &\overset{(i)}{\approx} -\frac{2\Omega^2\gamma_\alpha}{\Gamma^2} \qty[N_i - 2\frac{N_i \qty(N_g + 1)}{N_g + 1 + \frac{1}{C}} + \frac{N_i\qty(N_g+1)\qty(N_g + 1)}{\qty(N_g + 1 + \frac{1}{C})^2}] \nonumber \\
    &\overset{(ii)}{\approx} \frac{2\Omega^2\gamma_\alpha}{\Gamma^2} \frac{N_i}{\qty(N_g+1)^2C^2}\, . \label{eq:app_rates2}
\end{align}
Now we can justify assumption (i): For $N_gC \gg 1$ the cavity decay channel is dominant, for which $J$ does not change. As $J = N/2$ for the initial state, we can thus expect $J$ to remain close to this value. In fact, the dominant spontaneous emission rate $\gamma_x$ can only decrease $N - 2J$, pushing the system back into the superradiant state $J=N/2$ if it moves out of that state during the initial dynamics. This corresponds to the results of numerical simulation, where initially $N - 2J$ grows, but then quickly decays back towards zero.

Note that the Eqs.~\eqref{eq:app_rates1} to \eqref{eq:app_rates2} are still state dependent. In order to get rate equations we need to take the expectation value of the right hand side. For $N \gg 1$, the fluctuations of the particle numbers $N_i$ and $N_g$ around their mean values are typically small compared to their expectation values. Therefore, the expectation values of the right hand side are well approximated by taking the expectation value of $N_g$ directly in the denominator, recovering the rate equations given in the paper. Note that by comparing simulations of the rate equations to simulations of the full master equation, we find that up to a prefactor for the loss state population $N_x$ they give a good approximation to the dynamics.

\section{Solution of Rate Equations} \label{sec:app_rate_solution}

For large $N_gC$, we note that $N_x \ll N$, as $\dot N_x \ll \dot N_g$. Thus, to first order $N_g = N - N_i$. For large $N$, we can thus write down a differential equation for the initial state population:
\begin{align}
    \dot n_i \equiv \dot N_i / N \approx - \frac{2}{NC\tau} \frac{n_i}{1 - n_i} 
\end{align}
with $\tau = \Gamma/\Omega^2$
This equation can be integrated to find for the time $T$ to reach a population fraction $n_i$ in the initial state:
\begin{align}
    T(n_i) = \int_1^{n_i} \frac{\d n_i'}{\dot n_i'} = \frac{NC\tau}{2} \ln(\frac{e^{n_i-1}}{n_i})
\end{align}
or inverted to get the time evolution of $n_i$
\begin{align}
    n_i(t) = - W\qty[-\exp(-1-\frac{2t}{NC\tau})]
\end{align}
where $W(x)$ is the product logarithm or Lambert $W$ function, which is defined as the inverse of $x = we^w$.

\section{Rate Equations with Detuning} \label{sec:app_polaritons}

In order to compute rate equations for the chirped pulse, we replace $\Delta = \delta = \Delta_\mathrm{opt}$ with
\begin{align}
    \Delta_\mathrm{opt} &= \left\{\max\qty[0, \qty(N_g+1)g^2 - \frac{\Gamma^2 + \kappa^2}{2}]\right\}^{1/2} .
\end{align}
If we assume $(N_g+1)g^2 > (\kappa^2+\Gamma^2) / 2$, the cavity decay dominates and we can restrict the analysis to the completely symmetric Dicke state, for which one can replace $\hat S_{ig} \hat S_{gi} = \hat N_i (\hat N_g + 1) / N $. This leads to:
\begin{align}
    \hat L_\mathrm{eff}^{\kappa\dagger} \hat L_\mathrm{eff}^\kappa &= \Omega^2g^2\kappa \frac{\hat N_i \qty(\hat N_g + 1)}{\qty[ \qty(\hat N_g - N_g)g^2 - \frac{\qty(\kappa-\Gamma)^2}{2}]^2 + \qty(\kappa + \Gamma)^2 \qty[\qty(N_g+1)g^2 - \frac{\Gamma^2 + \kappa^2}{2}]} \nonumber \\
    &\approx \frac{\Omega^2 \kappa}{\qty(\Gamma + \kappa)^2}\hat N_i \\
    \hat L_\mathrm{eff}^{\alpha\dagger} \hat L_\mathrm{eff}^\alpha &= \frac{\Omega^2\gamma_\alpha}{\Gamma^2 + \qty(N_g+1)g^2} \qty{ \hat N_i - \frac{2\hat N_i \qty(\hat N_g + 1)g^2\kappa\Gamma + \hat N_i \qty(\hat N_g + 1)^2g^4}{\qty[ \qty(\hat N_g - N_g)g^2 - \frac{\qty(\kappa-\Gamma)^2}{2}]^2 + \qty(\kappa + \Gamma)^2 \qty[\qty(N_g+1)g^2 - \frac{\Gamma^2 + \kappa^2}{2}]} } \nonumber \\
    &\approx \frac{\Omega^2 \gamma_\alpha}{\qty(\Gamma + \kappa)^2}\hat N_i
\end{align}
where we approximated by replacing operators with their expectation values and neglecting higher order terms in $(\Gamma + \kappa) / \sqrt{N_g+1}g$. 

However, in numerical simulations the noise term $\hat N_g - N_g$ turns out to be important as well. Firstly, for $\Delta N_g \sim \sqrt{N_g}$, we find that the noise term looks like $N_g g^4$. This is only negligible if $N^{1/4}g \ll \kappa + \Gamma$. Secondly, we can end up in a negative feedback loop, running out of resonance: Consider a state for which $\langle \hat N_g \rangle(t_1) < N_g(t_1)$. In this case $\partial_t \langle \hat N_g \rangle \sim \langle \hat L_\mathrm{eff}^{\kappa\dagger} \hat L_\mathrm{eff}^\kappa \rangle < \partial_t N_g \sim \left.\langle \hat L_\mathrm{eff}^{\kappa\dagger} \hat L_\mathrm{eff}^\kappa \rangle \right\rvert_{\hat N_g = N_g} $. This leads to $(\langle \hat N_g \rangle - N_g)(t_2) < (\langle \hat N_g \rangle - N_g)(t_1)$ for $t_2>t_1$, running further out of resonance. In practice this can be solved by keeping $\Delta$ a bit smaller than $\Delta_\mathrm{opt}$. We choose
\begin{align}
    \Delta &= \left\{\max\qty[0, N\qty(\frac{N_g(t)}{N})^{1.5}g^2 - \frac{\Gamma^2 + \kappa^2}{2}]\right\}^{1/2}
\end{align}
with $N_g(t)/N$ plugged in according to an empiric estimate
\begin{align}
    \frac{N_g(t)}{N} &= \frac{\qty(\kappa + \gamma_g)\qty(\kappa + \gamma_x)}{\qty(\kappa + \gamma_g + \gamma_x)^2} \qty{1 - \exp[-\frac{2\Omega^2 \qty(\kappa + \gamma_g + \gamma_x)t}{\qty(\kappa + \Gamma)^2}]} \nonumber \\
    &\quad + \frac{\gamma_g^2}{\qty(\kappa+\gamma_g + \gamma_x)\qty(\gamma_g + \gamma_x)} \qty{1 - \exp[\frac{2\Omega^2\qty(\gamma_g+\gamma_x)t}{Ng^2-\frac{\Gamma^2}{2} + \frac{\kappa^2}{2}}]}
\end{align}
This choice yields very good results, and a further optimization is beyond the scope of this paper.

\section{Dark States} \label{sec:app_dark}

As discussed above and shown in Tab.~\ref{tab:dicke_master}, only spontaneous emission towards $\ket i$ and $\ket g$ can decrease $J - N/2$. This effect corresponds to a symmetry breaking, as only the $J=N/2$ states are completely symmetric at the state level. The resulting states with lowered $J$ can still decay via the cavity until $M = -J$, or reformulated $N_i = N/2 - J$. This however leaves $N/2 - J$ molecules trapped in the initial state, as $\hat L_\kappa^\mathrm{eff} \ket{N/2, J, M = -J} = 0$. The remaining states are dark states and will decay with a lower rate. Note that the dark states do not have shifted energies with respect to the bare excited states. Thus the laser is detuned with respect to the initial state -- dark state transition by about $\Delta^2 \sim Ng^2$. This leads to a further lowered decay rate.

We can estimate the number of molecules that are trapped in these dark states by assuming that every spontaneous emission towards $\ket g$ reduces $J$, whereas spontaneous emission towards $\ket i$ leaves $J$ unchanged. This rough approximation is empirically justified for $J \approx N/2$, and is consistent with the finding of Ref.~\cite{Gegg_2018_Superradiant} that spontaneous emission is more relevant for dark states than dephasing. With this assumption the dark state fraction becomes $\gamma_g / (\kappa + \gamma_g + \gamma_x)$. The decay rate of this dark states is given by $2\Omega^2(\gamma_g + \gamma_x) / (\Delta^2 + \Gamma^2) \approx 2\Omega^2(\gamma_g + \gamma_x) / (Ng^2 - \kappa^2/2 + \Gamma^2/2)$.

\section{Comparison between both Schemes} \label{sec:app_comp}

For given $g$, we find that the chirped pulse scheme gives a loss state population of:
\begin{align}
    \frac{N_x^\infty}{N}(\mathrm{chirp}) \approx \frac{\gamma_x}{\kappa} \gg \frac{\gamma_x}{\sqrt{N}g} \gg \frac{\gamma_x}{\sqrt{N}g} \frac{\kappa}{\sqrt{N}g} = \frac{f_x}{NC} \approx \frac{N_x^\infty}{N\ln(N)} (\mathrm{no\ chirp})
\end{align}
Thus for $\ln(N) < Ng^2 / \kappa^2$ the scheme without a chirp yields higher state selectivity.

In contrast, for given half time we find:
\begin{align}
    \frac{N_x^\infty}{N}(\mathrm{no\ chirp}) &\approx \frac{f_x\tau \ln(N)}{5T_\frac{1}{2}} \\
    T_\frac{1}{2} (\mathrm{chirp}) &\approx \frac{\ln(2) \kappa}{\Omega^2} \approx \frac{\ln(2)\tau\kappa}{\Gamma} \\
    \frac{N_x^\infty}{N}(\mathrm{chirp}) &\approx \frac{f_x \tau \ln(2)}{T_\frac{1}{2}}
\end{align}
Thus, for $N \gtrsim 30$, the state selectivity with the second scheme is higher. For a typical number of $N \sim 1000$ molecules we find that the state selectivity with the chirped pulse is higher by a factor of \num{2}. Note also that the first scheme does not decay exponentially, but the long time dynamics exhibits a stronger collective slowdown so that considering the total transfer time instead of the half time favors the chirped scheme even more.

\section{Influence of Local Cavity Coupling Constant} \label{sec:app_disorder}

In order to model local cavity coupling constants for large molecule numbers, we use an effective model. We define a threshold cooperativity $C_\mathrm{thr}$, and assume that molecules with local coupling $C_n = g_n^2/(\kappa\Gamma) < C_\mathrm{thr}$ do not couple to the cavity $g_n \rightarrow 0$, whereas molecules with $C_n > C_\mathrm{thr}$ couple with average cooperativity $C_n \rightarrow C_\mathrm{eff} = (\sum C_n) / N'$. By employing this binary decision model, we arrive at a situation with particle permutation symmetry, which can be simulated as described above. $C_\mathrm{thr} = C_\mathrm{thr}(N)$ is chosen such that, if $N$ molecules couple with $C_\mathrm{thr}$ to a cavity, the $\ket x$ state fraction is half of its no cavity value [red dashed line in Fig.~\ref{fig:disorder}(a)]. Errors are given by or smaller than the linewidth.

\begin{figure}
	\centering
	\includegraphics[width=\textwidth]{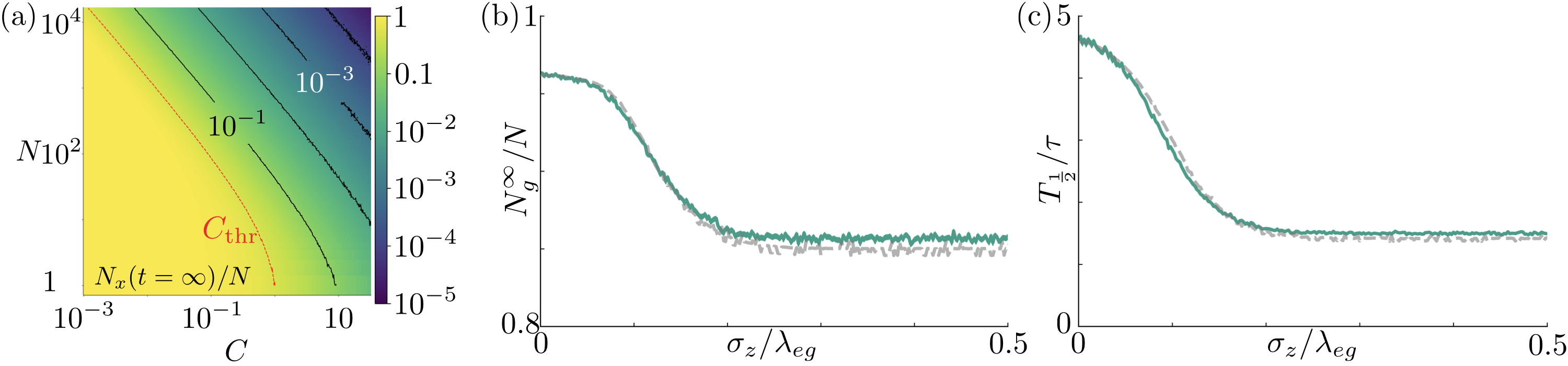}
	\caption{(a) Contour plot of $N_x^\infty / N$ with indicated choice of $C_\mathrm{thr}$ for the disorder model according to the red dashed line. This corresponds to $N_x^\infty$ being half of its no cavity value. (b, c) Comparison of (b) the final ground state population and (c) the halftime calculated with the full simulation (continuous line) and the effective model (dashed line). We choose $N=10$ and $C_0 \equiv g_0^2/(\kappa\Gamma)\approx4$. The local coupling strength is given by $g(z) = g_0\cos(2\pi z/\lambda_{eg})$ and $z$ is randomly distributed with probability density function $p(z) = \exp[-z^2 / (2\sigma_z^2)]$.}
	\label{fig:disorder}
\end{figure}

In order to derive an expression for $C_\mathrm{eff}$, we analyze the limit for which the cavity decay is dominant. We first derive an expression for the states after $k$ decay processes via the cavity. Then, we calculate the cavity decay rate and the spontaneous emission rates for these states. Comparing the rates of decay for the different decay channels we can estimate the final ground state population and the half time.

For a local cavity coupling constant, we find effective operators for the master equation after adiabatic elimination
\begin{align}
	\hat{H}_\mathrm{eff} &= 0 \\
	\hat L_\mathrm{eff}^\kappa 
	&= \sqrt{\lambda_\kappa} \hat{\xi} \hat{S}_{gi} \\
	\hat L_\mathrm{eff}^{{\alpha}, {(n)}} 
	&=  \sqrt{\lambda_\gamma^\alpha}  \left( \hat{\sigma}^{(n)}_{\alpha i} - \frac{\hat{\sigma}^{(n)}_{\alpha g}g_n}{g_0} \hat{\xi} \hat{S}_{gi} \right)
\end{align}
with $\hat S_{g i} = \sum_n g_n \hat \sigma_{g i}^{(n)} / (\sqrt{N}g_0) $ and $\hat \xi = \sqrt{N} g_0^2 (\sum_n g_n^2 \hat \sigma_{gg}^{(n)} + \kappa\Gamma)^{-1}$, where $g_0$ is the peak cavity coupling constant, $\bar C$ is the average cooperativity and $C$ is the peak cooperativity.

Taking the initial state as $\ket{\psi^{(0)} } = \bigotimes_n \ket i_n$, the state after $k$ decay processes via the cavity decay channel $\hat L_\mathrm{eff}^\kappa$ is given by 
\begin{align}
    \ket{\psi^{(k)}} = \frac{1}{\mathcal N_k} \qty(\hat L_\mathrm{eff}^\kappa)^k \ket{\psi^{(0)}} = \frac{1}{\mathcal N_k'} \sum_{I, G} \bigotimes_{l \in I} g_l \ket i_l \otimes \bigotimes_{m \in G} \ket{g}_m
\end{align}
for some normalization constants $\mathcal N_k$ and $\mathcal N_k'$. $I$ denotes the set of atom pairs in state $\ket i$ and $G$ denotes the set of molecules in state $\ket g$. The sum runs over all possible choices of sets $I$ and $G$ such that $|G| = k$, $|I| = N - k$, and $I \cap G = \emptyset$. This can be easily confirmed by checking $\ket{\psi^{(0)}} = \bigotimes_n \ket i_n$ and $\hat L^\kappa_\mathrm{eff} \ket{\psi^{(k)}} \propto \ket{\psi^{(k+1)}}$. In these states we get to leading order in $1/(N_g\bar C)$:
\begin{align}
    \zeta_\kappa &\approx \frac{2}{\tau} \frac{N_i}{(N_g + 1)\bar C} \\
    \zeta_\alpha &\approx \frac{2f_\alpha}{\tau} \frac{N_i}{(N_g+1)^2 \bar C^2}\, .
\end{align}
This also justifies in hindsight to look at the cavity dominated limit, as cavity decay dominates for $N_g\bar C \gg f_\alpha$.
From these we get the equations of motion:
\begin{align}
    \dot N_i &\approx - \frac{2}{\tau} \frac{N_i}{(N_g + 1)\bar C} \label{eq:eom_dis_i}\\
    \dot N_g &\approx \frac{2}{\tau} \frac{N_i}{(N_g + 1)\bar C} \label{eq:eom_dis_g}\\
    \dot N_x &\approx \frac{2f_x}{\tau} \frac{N_i}{(N_g+1)^2 \bar C^2} \label{eq:eom_dis_x}
\end{align}
These equations are equivalent to Eqs.~\eqref{eq:cav_dec} and \eqref{eq:spon_dec} and the rate equations $\dot N_{i,g,x}$ in the main paper for no disorder with the replacement $C \rightarrow \bar C$. This motivates the choice:
\begin{align}
    C_\mathrm{eff} = \frac{\sum_n C_n}{N'}
\end{align}
where the sum over molecules with $C_n > C_\mathrm{thr}$ and $N'$ is the number of these molecules.

Fig.~\ref{fig:disorder}(b) and (c) show a comparison of the effective model and a full simulation for $10$ molecules and different disorder strengths. We find a good correspondence.

\section{Finite Temperature Effects}\label{sec:app_finT}

\begin{figure}
    \centering
    \includegraphics[width=0.9\textwidth]{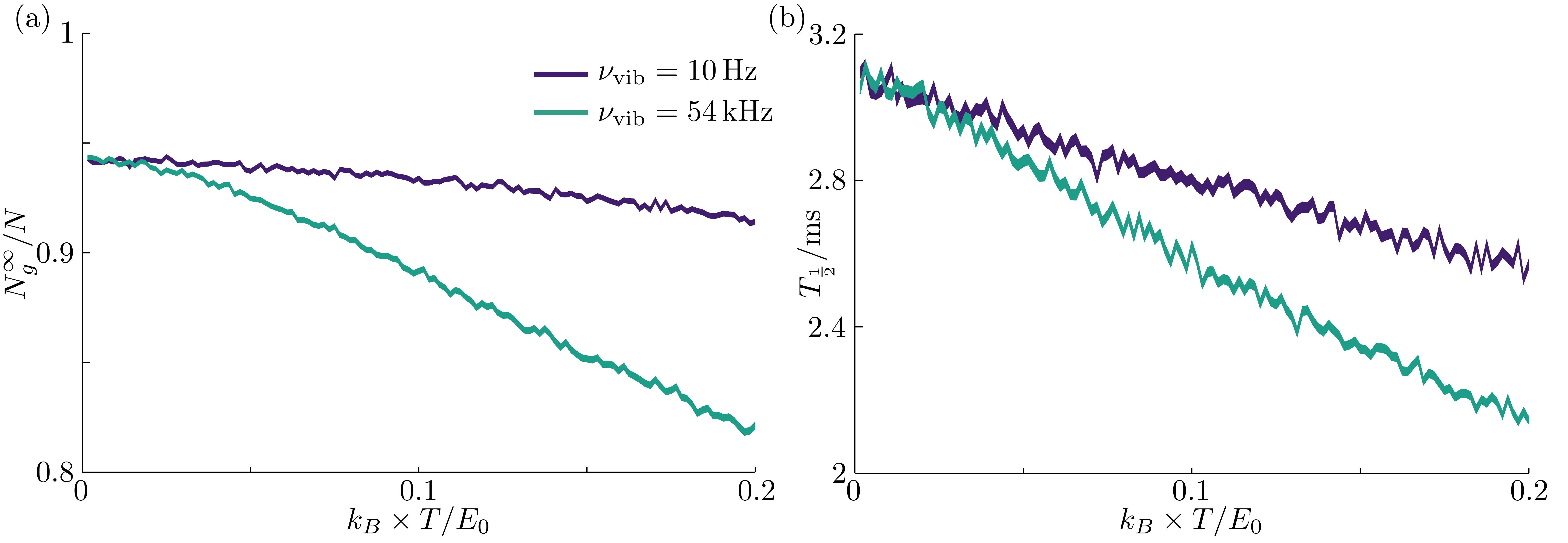}
    \caption{Temperature dependence of (a) the final molecular fraction and (b) the transfer half time. We simulate $N=6$ molecules, that oscillate classically at frequency $\nu_\mathrm{vib}$, for a peak cavity coupling constant $g_0/2\pi = \SI{10}{\mega\hertz}$, with thermal energy up to \SI{20}{\percent} of the lattice depth. The linewidth corresponds to the statistical error due to the trajectories and the choice of $z_{0,i}$ and $\phi_{0,i}$ (one standard deviation). The violet and green line correspond to $T_\frac{1}{2}\nu_\mathrm{vib}<1$ and $T_\frac{1}{2}\nu_\mathrm{vib}>1$, respectively.}
    \label{fig:tscaling}
\end{figure}

Our scheme is intended to work in the ultracold regime, where temperatures $T$ are smaller than the inter-band lattice energy gap $h \nu_{\mathrm{vib}}$ and the effects of finite $T$ are negligible.  For typical lattice depths, this entails temperatures of several hundreds of \si{\nano\kelvin} (see below), which is achievable in many experiments. We note that it is commonly achieved to prepare a Mott insulator state in the lowest lattice band~\cite{Rom_State_2004,Covey_Controlling_2017}. However, the precise distribution of molecules in the lattice will depend on the chosen preparation and loading scheme. 
In the following we assume that molecules in any doubly occupied site are rapidly lost via three-body recombination or light-induced collisions in the early stages of our scheme. This leaves a lattice with at most one Feshbach molecule per site. In the following, we estimate the efficiency of our scheme when population of higher bands cannot be neglected, i.e., for $k_B T\gtrsim h \nu_{\mathrm{vib}}$ with $k_B$ the Boltzmann constant.\\

 Due to the unfavorable scaling of the size of the density matrix with increasing $N$ and band number, a computation of the quantum many-body dynamics becomes impractical for just a few particles. We thus estimate the efficiency of the scheme in a classical approximation for the on-site motion of the molecules. This corresponds to treating the molecules in coherent states, which, compared to typical experimental scenarios, largely overestimates the number of molecules in higher bands. We distinguish two regimes depending on how fast the population transfer $T_{\frac{1}{2}}$ occurs compared to $1/\nu_{\mathrm{vib}}$, i.e.\ the characteristic oscillation time of a particle at the bottom of a lattice well. We find that if $T_{\frac{1}{2}}\nu_\mathrm{vib}<1$, the system is well modelled by adding static disorder to the cavity coupling constant (due to positional disorder inside the lattice). In this case, we find that the efficiency of the scheme is only slightly decreased from the zero-temperature case  [see Fig.~\ref{fig:tscaling}, upper violet curves]. If instead $T_{\frac{1}{2}}\nu_\mathrm {vib}\gtrsim 1$, as for the parameters in the paper, the decrease in efficiency is larger. However, we estimate that reasonably good transfer rates are still possible, as discussed below [see also Fig.~\ref{fig:tscaling}, lower green curves]. We note that the decrease in efficiency is always accompanied by a speed-up in the transfer dynamics.
Finally, if the thermal energy is on the order of the lattice depth $E_0$, molecules are lost from the trap, leading to a steep decrease in transfer efficiency. In the following we provide more details on our calculations.\\

The lattice band spacing can be calculated from the potential and its harmonic approximation:
\begin{align}
    V(\vec x) &= E_0 \sin^2\qty(\frac{2\pi x}{\lambda_\mathrm{latt}}) +
    E_0 \sin^2\qty[\frac{2\pi(\cos\theta z + \sin \theta y)}{\lambda_\mathrm{latt}}] +
    E_0 \sin^2\qty[\frac{2\pi(\cos\theta z - \sin \theta y)}{\lambda_\mathrm{latt}}] \\
    &\approx \frac{4\pi^2 E_0}{\lambda_\mathrm{latt}^2} \qty[x^2 + \qty(2\sin^2\theta) y^2 + \qty(2\cos^2 \theta) z^2] \label{eq:harm_pot}
\end{align}
with the parameters of the main text $\lambda_\mathrm{latt} = \SI{1064.5}{\nano\metre}$, $\theta=\SI{57}{\degree}$, and $E_0 = 48E_R = k_B \times \SI{1.8}{\micro\kelvin}$ for the recoil energy $E_R = (2\pi\hbar/\lambda_\mathrm{latt})^2/(2m_\mathrm{RbCs})$. We will restrict this analysis to motion along the $z$-direction, as here the variation of the cavity coupling constant with motion is largest and consider the one dimensional potential $V(z)=V(x=y=0,z)$. The oscillation frequency in the $z$-direction is given by $\nu_\mathrm{vib} = (2 \cos \theta/\lambda_\mathrm{latt})\sqrt{E_0/m_\mathrm{RbCs}} = \SI{54}{\kilo\hertz}$.
We model the thermal motion by $z_i(t) = z_{0,i}\cos(2\pi\nu_\mathrm{vib} t + \phi_{0,i})$, with Boltzmann distributed amplitudes $z_{0,i}$ chosen according to $p(z_{0,i}) = \exp[-V(z_{0,i})/(k_BT)]/Z$ with the partition function $Z=\int \d z \exp[-V(z)/(k_BT)]$, and a random initial phase $\phi_{0,i} \in (0,2\pi]$. This thermal motion leads to a time dependent cavity coupling constant for each molecule $g_i(z_i) = g_0 \cos[2\pi z_i/\lambda_{eg}]$, which is incorporated in the simulation. We choose $g_0/2\pi=\SI{10}{\mega\hertz}$ to get $T_\frac{1}{2}\simeq\SI{3}{\milli\second}$ 
in order to approximate the transfer half time $T_\frac{1}{2}$ computed in the main text for RbCs. This allows us to provide results for characteristic transfer half times $T_\frac{1}{2}\nu_\mathrm{vib}\gtrsim 1$, similar to the large-$N$ case of the main text, however for the case of just a few particles. The results are presented in Fig.~\ref{fig:tscaling}, which shows a moderate decrease of both transfer efficiency and transfer time. Since the key parameter is here the quantity $T_\frac{1}{2}\nu_\mathrm{vib}$, we expect that similar results should hold also for larger $N$.

All temperature induced frequency shifts can be ignored, as both shifts of thermally excited states compared to the ground state $\Delta_\nu \sim k_BT/ h$ as well as Doppler broadening $\sigma_D = \sqrt{k_BT/(m_\mathrm{RbCs}\lambda^2)} \sim \si{\kilo\hertz}$ are on the order of \si{\kilo\hertz}, much smaller than the natural linewidth $\Gamma$ or cavity linewidth $\kappa$, which are on the order of \si{\mega\hertz} or \si{\giga\hertz}, respectively.

\end{widetext}

\end{document}